\documentstyle[preprint,aps,epsf]{revtex}
\tightenlines
\begin{document}
\draft

\title{Fast Tree Search for Enumeration of a Lattice Model of Protein Folding}
\author{Henry Cejtin\thanks{Present address: Sourcelight Technologies Inc.,
906 University Place, Suite B-211, Evanston, IL 60201}, 
Jan Edler\thanks{Present address: Yianilos Lab, 707 State Road, 
Suite 212, Princeton, NJ 08540}, 
Allan Gottlieb\thanks{Also at: Computer Science Department, NYU, New
York, NY 10003}, 
Robert Helling\thanks{Present address: Institut fuer Physik,
Humboldt-Universitaet zu Berlin, Berlin, Germany}, 
Hao Li\thanks{Present address: Department of Biochemistry and Biophysics,
University of California at San Francisco, San Francisco, CA 94143},
James Philbin\thanks{Present address: StorageNetworks, Inc.,
4 Independence Way, Princeton, NJ 08540},
Chao Tang\thanks{Author to whom correspondence should be addressed;
electronic mail: tang@research.nj.nec.com}
and Ned Wingreen} 

\address{NEC Research Institute, 4 Independence Way, Princeton, New
Jersey 08540}

\date{\today}
\maketitle

\begin{abstract}

Using a fast tree-searching algorithm and a Pentium cluster, we
enumerated all the sequences and compact conformations (structures) for
a protein folding model on a cubic lattice of size $4\times3\times3$. 
We used two types of amino acids -- hydrophobic (H) and polar (P) -- to
make up the sequences, so there were $2^{36} \approx 6.87 \times 10^{10}$
different sequences. The total number of distinct structures was $84,731,192$. 
We made use of a simple solvation model in which the energy of a sequence 
folded into a structure is minus the number of hydrophobic amino acids in
the ``core'' of the structure. For every sequence, we found its ground state
or ground states, {\it i.e.}, the structure or structures for which its 
energy is lowest. About $0.3\%$ of the sequences have a unique ground state.
The number of structures that are unique ground states of at least one
sequence is $2,662,050$, about $3\%$ of the total number of structures.
However, these ``designable'' structures differ drastically in their
designability, defined as the number of sequences whose unique ground state
is that structure. To understand this variation in designability, we
studied the distribution of structures in a high dimensional space in which
each structure is represented by a string of 1's and 0's, denoting core and
surface sites, respectively.

\end{abstract}

\newpage
\narrowtext

\section{INTRODUCTION}

The protein folding problem \cite{cre} has long attracted the attention
of scientists from various disciplines. The relationship between the 
amino-acid sequence and the three-dimensional structure of a protein
is not only an
extremely important and practical problem in biology, but also a
fundamental problem in science. Despite a tremendous amount of effort and
progress over many decades, the problem remains essentially unsolved. At
least part of the difficulty arises from the intrinsic complexity of the
protein-folding problem. Since the seminal work of 
Anfinsen \cite{anf} about 40
years ago, it has been demonstrated that the native state of a small,
single domain protein is the global minimum of the free energy. However,
the minimum-free-energy conformation of a polypeptide chain is ``hiding'' in
a large space of $z^N$ conformations, where $N$ is the length of the
chain and $z$ is the effective coordination number. Even if we count only
the compact conformations for which $z \approx 2$ (see below) and take
$N=100$ for typical small proteins, the number of conformations is huge,
$z^N \approx 2^{100} \sim 10^{30}$.
On top of this huge conformational space lies the heterogeneity of 
amino acids. The 20 natural amino acids differ in size, hydrophobicity, and
other physical and chemical properties. This heterogeneity is coupled
with two intrinsic features of polymers: chain connectivity and the excluded
volume effect. The free energy is a sensitive and complicated
function in this huge conformational space with complex constraints. 

\begin{figure}
%\vspace{4cm}
\centerline{\epsfxsize=8.5cm
\epsffile{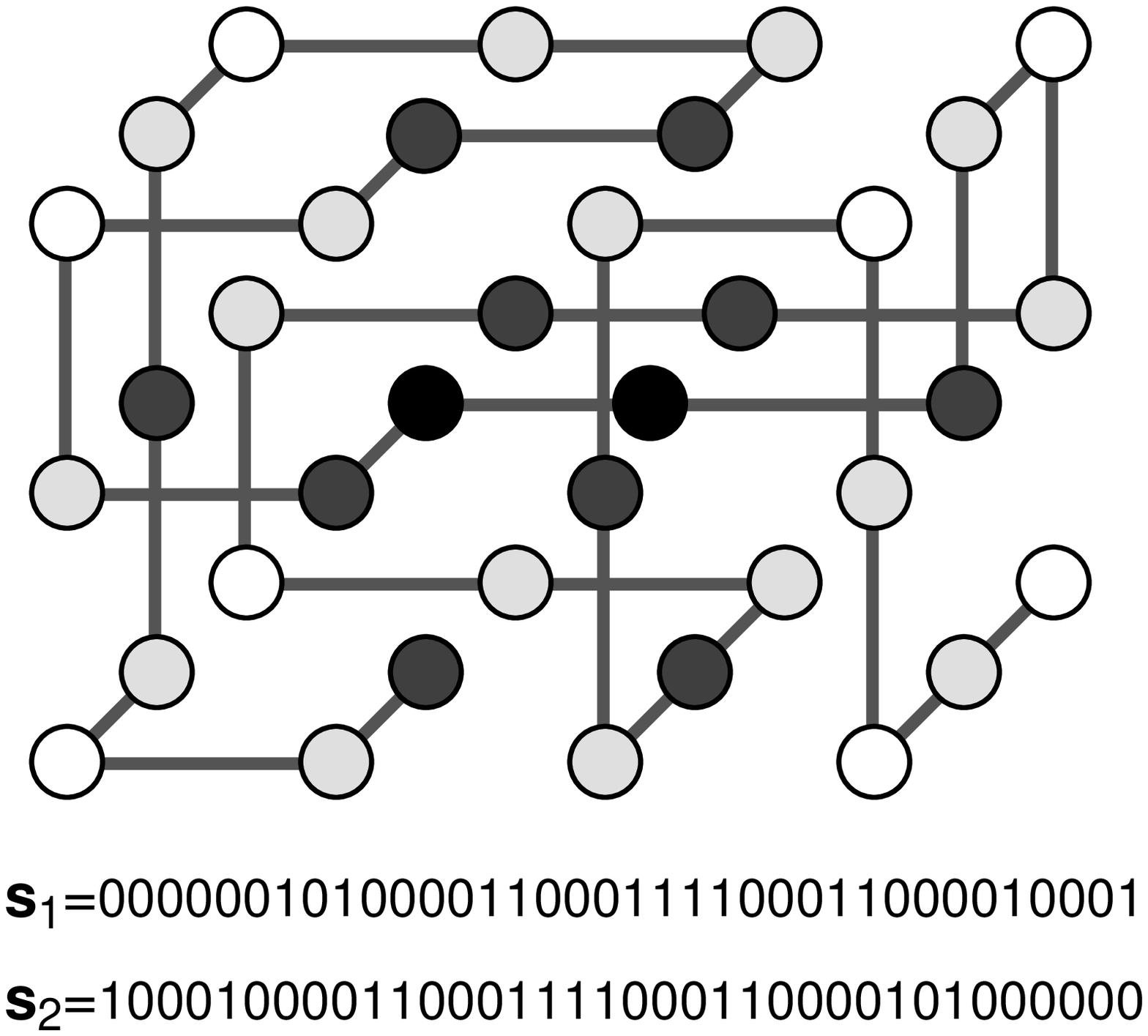}}
\vspace{1cm}
\caption{A compact $4 \times 3 \times 3$ conformation on a cubic
lattice. The sites are classified into centers (black), faces (dark gray),
edges (light gray), and corners (white). This geometrical conformation
corresponds to two structures, one starting with each end of the chain.
Shown below the conformation are its corresponding two binary strings, in
which 1's correspond to ``core" sites (centers and faces) and 0's correspond
to ``surface" sites (edges and corners).}
\label{lattice}
\end{figure}

In the last decade or so, there has been increasing interest in studying
simple lattice models of protein folding. 
In these models, polypeptide chains are represented by self-avoiding
walks on a regular lattice ({\it e.g.}, Fig.~\ref{lattice}), 
greatly simplifying the conformational space. 
Very often the sequence space and hence the
heterogeneity is also simplified by using only two types of amino
acids: hydrophobic (H) and polar (P). These so-called ``HP lattice
models'' \cite{dill85,lau89,chan93} nonetheless capture some essential
features of the protein
folding problem. Simple lattice models have been applied to a wide
range of problems including collapse and folding transitions
\cite{sali94,socci94,shriva95,klimov96,melin99},
the influence of packing on secondary-structure formation \cite{chan91}, and
differences in the designability of structures \cite{gov95,li96}.
The advantage of HP lattice models is that they are
simple enough to be amenable to thorough theoretical study. These
studies can provide fruitful insights to feed back to 
or test against realistic models and experiments.

One approach for calculating thermodynamic and other properties
of lattice models is to enumerate all possible 
sequences and conformations \cite{lau89,chan93,li96,shak90,cam93,pande94}. 
Since native globular
protein structures and presumably most of the low-energy states are
compact, enumeration studies are usually done on compact conformations
only. The number of compact conformations $C_N$ scales with the chain
length $N$ as $C_N \sim z^N$, where $z$ is an effective coordination number.
Numerical estimation gives $z \approx 1.47$ for 2D square lattices
\cite{cam93,sch} and $z \approx 1.86$ for 3D cubic lattices \cite{pande94}, in
good agreement with mean-field calculations of $Z/e$ \cite{orl} and
$(Z-1)/e$ \cite{flo}, respectively, where $Z$ is the coordination number
of the lattice and $e=2.718\cdots$ is the base of the natural logarithm.
(For real 
peptide chains the number of ``distinct'' states an amino acid can take, as
estimated very roughly from Ramachandran plots of dihedral-angle 
frequencies, is about 5 or 6,
which gives $z \approx 2$.) 
For HP models in which there are only two types of amino acids, the
number of sequences is $2^N$. If in the enumeration study the energies
of every sequence folded into every compact conformation are evaluated,
the total number of energy calculations is $2^N \times C_N$. The largest
system previously evaluated in this way 
is an $N = 27$ ($3 \times 3 \times 3$) cubic
lattice model \cite{li96} where $2^N \times C_N = 2^{27} \times 103346
\approx 10^{13}$.
Several interesting results were found in the enumeration of the
27-mer, in particular the idea of designability and its relation to 
thermodynamic stability \cite{li96}. However, $N=27$ is still small
compared with typical protein sizes. It would be very desirable to
enumerate larger systems if at all possible. 
In this paper, we report results of a complete enumeration of an $N = 36$
($4 \times 3 \times 3$) cubic-lattice model. The number of compact
conformations is $84,731,192$ \cite{pande94}, so, naively, the total number
of energy calculations is $2^{36} \times 84,731,192 \approx
6\times10^{18}$. The task was made possible by using a binary model,
a fast tree-search algorithm yielding a speed-up factor of 1600, 
and a 53-processor 200MHz Pentium Pro cluster.

\section{THE MODEL}

The protein folding model we use in the enumeration study is the
solvation model discussed in Ref.~\cite{li98}: Denote a sequence
of amino acids by $\{\sigma_i\}$. We take only two types of amino
acids: hydrophobic H ($\sigma=1$) and polar P ($\sigma=0$). 
A ``structure" is the set of all reflections and rotations
of a given compact conformation.
The energy of a sequence folded into a structure is taken
to be the sum of the contributions from each amino acid upon burial away
from water:
\begin{equation}
E = - \sum_{i=1}^N \sigma_i s_i,
\label{sol}
\end{equation}
where $s_i$ is a structure-dependent number characterizing the degree of
burial of the $i$-th amino acid in the chain. Larger $s_i$ corresponds to
smaller surface area accessible to solvent. For a 
$4 \times 3 \times 3$ structure on a cubic lattice
there are 4 different kinds of sites: center,
face, edge, and corner (see Fig.~\ref{lattice}). So in principle there
could be 4 different values of $s_i$. To simplify the calculation, we take
only two values for $s_i$: we define a string $\{s_i\}$ for each structure with
$s_i = 1$ if the $i$-th site is a ``core" (center or face) 
and $s_i = 0$ if it is a ``surface" (edge or corner). 
Thus each compact structure of $4 \times 3 \times 3$ is
mapped into a string of 1's and 0's, $\{s_i\}$, with 12 1's (``cores")
and 24 0's (``surfaces''). The surface-to-core ratio is 2, close to the
values for small natural proteins. 
With this simplification, the energy in Eq.~(\ref{sol}) is
just minus the dot product of two binary strings. For a given sequence
$\{\sigma_i\}$, a ground-state structure is one that minimizes
Eq.~(\ref{sol}). A sequence may have more than one ground state
structure, but we will be primarily interested in sequences with
unique ground states. Out of the $84,731,192$ compact structures,
the number of distinct structure strings is $14,062,236$,
among which $2,662,050$ (corresponding to $1,331,025$ lattice
conformations) each represent exactly one 
structure. Each of the remaining $11,400,186$ 36-bit strings represents
multiple structures. We also analyzed a $3\times3\times3$ model with 7
``cores'' (1 center and 6 faces) and 20 ``surfaces'' (12 edges and 8
corners). In this case, there are $103,346$ structures and $6,291$
distinct structure strings, among which only $120$ (corresponding to 60
lattice conformations) represent exactly one structure apiece.

\section{Tree Searching Algorithm}

Because the protein chains in our model are considered to be directed,
there are generally two structures for each geometrical conformation,
related by reversal of the direction of the chain (see Fig.~\ref{lattice}).
A small subset of structures are their own reversals. A structure string
can be its own reversal even if its associated structure is {\it not}
reversal symmetric. We found that among the
$14,062,236$ structure strings in the $4\times3\times3$ model, there are
$2,850$ which are their own reversals. The remaining $14,059,386$ strings
form $7,029,693$ pairs, with the two members of each pair being the
reversals of each other. To reduce memory use, we keep only one
member of each such pair, with an extra bit tagged on the string to indicate
that it actually represents two strings: itself and its reversal. There
are thus $7,029,693+2,850=7,032,543$ distinct strings which we
keep in the calculation.

Each of the $7,032,543$ distinct 36-bit structure
strings $\{s_i\}$ has exactly 12 1's and 24 0's. 
Our goal is to find a way, given a 36-bit sequence 
string $\{\sigma_i\}$, to find if there is a unique entry in
the table which maximizes the dot product of the structure
string with $\{\sigma_i\}$, or, equivalently, which minimizes the
energy of the sequence $\{\sigma_i\}$ according to Eq.~(\ref{sol}).

\begin{figure}
%\vspace{6cm}
\centerline{\epsfxsize=8.5cm
\epsffile{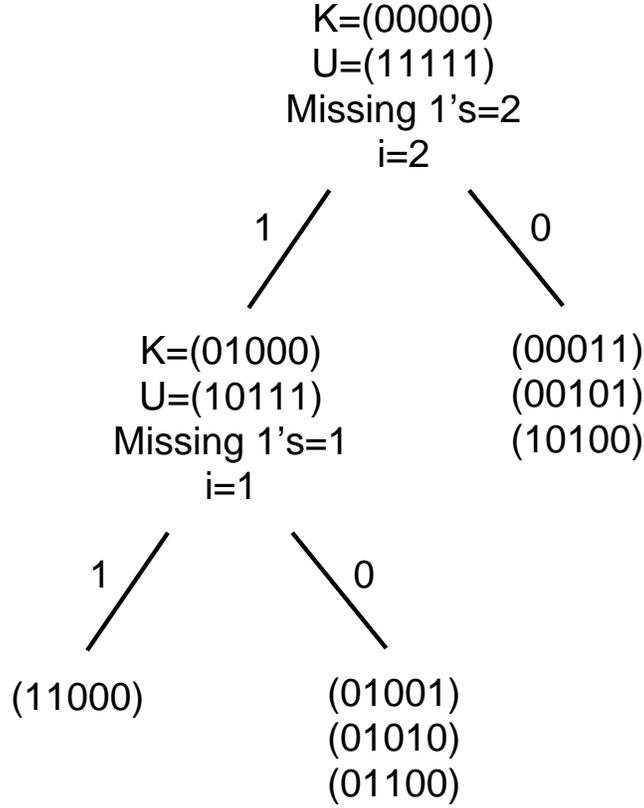}}
\vspace{2cm}
\caption{Example of a binary tree of structure strings.
The tree shown is constructed for seven strings of five bits        
each with exactly two 1's per string. By fiat, splitting stops      
and a leaf node is defined whenever the number of strings is three  
or fewer. At each node, $K$ gives the string ``Known-ones" and      
$U$ gives the string ``Undecided". Also indicated are the number    
``Missing ones" and the position $i$ on which the node branches.    
The tree used to search structure strings in the $4\times3\times3$
lattice HP model is simply a larger version of the tree shown here.}
\label{tree}
\end{figure}

To do this rapidly, we organized the strings in the table into a
binary tree (Fig.~\ref{tree}). 
First we describe how the tree is organized, and then
later how the tree was actually constructed. 

Each node in the tree represents a subset of the 7 million strings
in the table. For each node, the following information is
maintained:
\begin{itemize}
\item[(i)] Known-ones: A 36-bit string which has a 1 at the $i$'th 
position if and only if all the table entries corresponding to this 
node have 1's at the $i$'th position.
\item[(ii)] Undecided: A 36-bit string which has a 1 at the $i$'th 
position if and only if there is a table entry in this node which has 
a 1 at the i'th position and there is another table entry in this node 
which has a 0 at the $i$'th position.
\item[(iii)] Missing-ones: Each string in a node will have 1's at some
undecided positions. For each string, Missing-ones is the sum of these
1's. By construction each string has exactly 12 1's, so Missing-ones is
equal to 12 minus the sum of Known-ones. That is, Missing-ones is a
single integer no greater than 12 for each node.
\end{itemize}

If the node is not a leaf, then it also contains a position number $i$ and two
child nodes. These children partition the entries in the parent according
to the value of the indicated position $i$: one child has all the 
parent entries where $i=1$ and the other child has all of
the entries where $i=0$. Each leaf node at the end
of the tree contains a small list of structure strings -- in practice
we found 16 strings per leaf to work best. 

Given a 36-bit sequence string $\{\sigma_i\}$ and a node of the tree, 
what bounds can we place on the dot product of $\{\sigma_i\}$ and all
structure strings represented by the node?
Clearly, for all strings in the node the total dot product is
at least as big as the dot product of $\{\sigma_i\}$ with 
Known-ones. On the other hand, the total dot product is at most
$\{\sigma_i\}$ dotted with Known-ones plus the dot product of
$\{\sigma_i\}$ with Undecided. 
Another upper bound for the total dot product is $\{\sigma_i\}$
dotted with Known-ones plus the integer Missing-ones. 

Given such a tree, where the root corresponds to all of the 7,032,543
entries, and a 36-bit sequence string $\{\sigma_i\}$, 
here is how we search the table:

    Compute an upper bound for the total dot product using the 
    smaller of the two upper bounds described above. Call this the
    ``goal" $G$.
    If there are any entries which achieve this goal, we are done.
    If not, repeat with the reversed version of the sequence. (This is
    necessary because our table contains only one member of each reversed
    pairs of structure strings. Whenever a ground state structure string
    is found for a sequence, the reversed sequence necessarily has the
    reversed structure string as a ground state.) Again, it we achieve
    the goal $G$, we are done.
    If not, decrease the goal $G$ by 1 and try again. Repeat
    until the goal is satisfied.

Given a goal $G$, we search the tree as follows, 
starting at the root node:

    If the bound on the node indicates that goal $G$ is unachievable,
return failure.

    If the node is a leaf node, check each entry.  If one is found
that satisfies the goal, return success. If not, then return failure.

    If the node is not a leaf node, try each of the children. We first try
the child that matches $\{\sigma_i\}$. That is, if the children split
on the value at the $j$'th position, then we refer to bit $j$ of 
$\{\sigma_i\}$. If it is a 1, then we first do the child having
all 1's at position $j$, and
second do the child having all 0's at position $j$.  Similarly,
if bit $j$ of $\{\sigma_i\}$ is a 0, then we first do the child having
all 0's at position $j$.

The essential advantage of the tree structure is that, typically,
we do not have to check many structure strings for each sequence
because nodes high up in the tree get eliminated by the 
upper bound. An additional advantage accrues because we are only
interested in sequences with unique ground states. Therefore, as
soon as two strings are found that satisfy the goal $G$, the 
search can be stopped for that sequence. Our protocol of following
the branches that match the sequence $\{\sigma_i\}$ is intended
to quickly identify strings which satisfy $G$.

We now discuss how the tree is actually built. In order to
take advantage of the natural clustering of structure strings,
we choose to split each node at the position that makes
each of the two child nodes as tightly clustered as possible.  
We measure this clustering for each child as the sum of the
``entropies'' for each bit, with the tightest clustering
corresponding to the minimum entropy. Specifically, for
each child we evaluate the total entropy $S$ of its set of 
structure strings as 
\begin{equation}
S = -N_{\rm child} \sum_i (p_i \ln p_i + q_i \ln q_i),
\label{entropy}
\end{equation}
where $p_i$ is the probability of
the $i$th position being a 1, and $q_i$ is the probability of
the $i$th position being 0, averaged over all $N_{\rm child}$
structure strings in the child node. We choose to split each node
at the position that minimizes the combined entropy of 
the two children.

The only remaining decision is when to stop splitting.  
When a node contains sufficiently few entries, 
it is fastest to just examine each entry in the node.
Different stopping sizes were tried, and 16 seemed to be optimal
to minimize search time per sequence. 
Another consideration was memory.
The memory used by the entries themselves is
        7,032,543 * 5 = 35,162,715 bytes = 33.5 megabytes,
but each node takes additional space.  With the given stopping criterion,
there were 1,384,679 nodes in the tree, and the total space required was
70.5 megabytes.

Constructing a tree in this manner took a few hours on a 200 MHz
Pentium Pro. Once the tree was constructed, carrying out each search
took, on average, about 800 microseconds per sequence, but this time 
varied widely from sequence to sequence.

For comparison, we also implemented a naive search algorithm
in which the energy of a sequence is computed for each structure
string. The tree-search algorithm ran approximately 1600 times
faster than our best variant of the naive approach.

\section{COMPUTING ALL THE GROUND STATES}

The tree-search algorithm allows us to quickly 
compute ground states for each protein sequence in turn,
and to record those sequences with unique
ground states, together with the corresponding structure string and
energy value.
When all this is done, we would also like additional statistics,
such as the designability of each structure, {\it i.e.}, for how 
many sequences it is the unique ground state.
This and other statistics can be computed, after the fact,
if the unique ground-state solutions are stored.

Because our protein chains are directed, {\it i.e.} the two
ends are not considered identical, both a structure and its
oppositely directed partner are allowed. (Sometimes these
are the same structure.) As a result, if a particular 
sequence $\{\sigma_i\}$ has a particular structure as a 
unique ground state, then the reversed sequence must have 
the reversed structure as its unique ground state. 
For the $4\times 3\times 3$ problem, there are therefore 
$(2^{35})+(2^{17})=34,359,869,440$ possible sequences
that need to be considered, counting all 36-bit binary strings
but rejecting reversed strings. 
There are 7,032,543 distinct structure strings. 
Since we are interested in {\it unique} ground states,
each structure string is tagged with an additional bit to indicate
whether it represents exactly one, or more than one, geometrical
structure.

The overall computation is trivially parallelizable
because calculating the ground state for each sequence can
be done independently.
In order to manage the calculation of ground states for all sequences
in parallel, it proved useful to divide the space of sequences into
``bundles", and use these as the unit of parallelism.
Instead of organizing these bundles by fixing some high-order bits
of the 36-bit sequence $\{\sigma_i\}$ and varying the rest,
we fixed an equal number of low- and high-order bits,
varying the bits in the middle.
This way entire bundles could be eliminated as 
reversals of the sequences in another bundle.
We performed our computations with 14 bits fixed and 22 bits varying,
which produced a total of $(2^{13})+(2^6)=8,256$ bundles.

These 8256 bundles were executed on the NECI Large Array
Multiple Processors (LAMP) system, which is a collection of 28 computers,
all containing 200MHz Intel Pentium Pro microprocessors.
Three machines were uniprocessors, the rest had two processors each.
Every machine ran the Linux operating system and had at least 128 MB of
memory.
Each computation typically required about 10 MB of memory for itself,
plus about 70 MB for the (read-only) tree of protein structure
information, which was shared in memory on the multiprocessors.
The dual processors handled two independent bundles concurrently,
and sharing the tree was necessary to keep the total memory ``footprint"
of the jobs small enough to fit together without conflict.

The distribution of bundles to ``workers" was handled by a single
``master'' machine running shell and AWK scripts to poll the others,
start new bundles, collect results, and detect any crashes that might occur.
The scripts were written to be restartable with minimal lost effort
in the event of a failure affecting their own operation.
As each bundle was completed, a compactly coded binary output file was
produced.
At the end, all 8,256 output files were merged into a single 450 MB
result file.
Auxiliary programs were written to extract human-readable
data from these binary files.

The complete computation ran for about 198 hours,
with an average of 39 processors running at any one time, giving a total
of 7,805 CPU hours.
Bundle execution times varied from 23 seconds to almost 19 hours,
with a mean of 56.7 minutes. 

\section{Results}

Using the tree-searching algorithm and the LAMP system, we were able to
completely enumerate the $4\times3\times3$ HP lattice protein model. 
We found that $114,572,949$
sequences have unique ground states, which is about $0.3\%$ of all
sequences. In comparison, $0.09\%$ of the sequences in the 
$3\times3\times3$ model have unique ground states. 

\begin{figure}
%\vspace{6cm}
\centerline{\epsfxsize=8.5cm
\epsffile{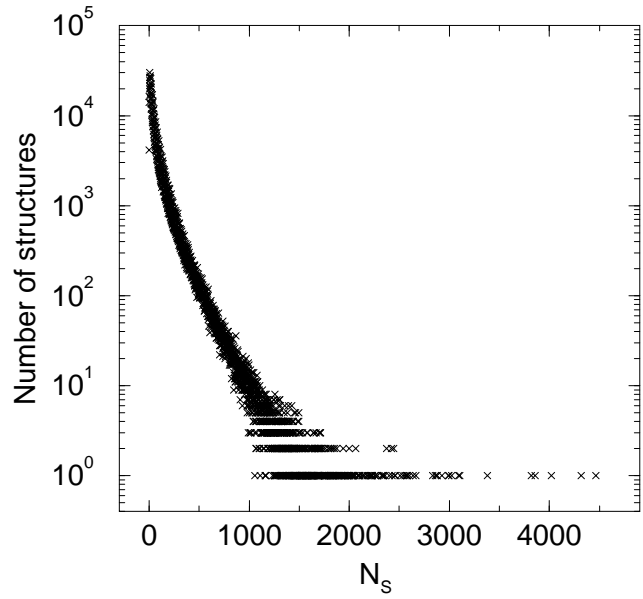}}
%\vspace {10cm}
\caption{Histogram for the designability of structures $N_S$.}
\label{histo}
\end{figure}

We associate to each
structure a quantity called designability \cite{li96,li98}. The
designability of a structure is the number of sequences having that 
structure as the unique ground state. By this definition, if two
or more structures share the same string representation then they
have zero designability since those structures can never be the
unique ground state of any sequence. Thus, only the $1,331,025$
structures (and their reverse paths) each of which has its own,
unshared string representation can have nonzero designability. 
For these structures, the average designability is
$114,572,949/1,331,025=86$. However, the designability of these
structures has a very broad range: from $1$ to $4,466$. 
(The minimum designability is 1 because the sequence with the
same bit-string as the structure is guaranteed to have that
structure as a unique ground state.) In Fig.~\ref{histo}, we
plot the number of structures with a given designability versus the
designability. One sees a long tail in the high designability region,
consistent with previous results on the $3\times3\times3$ model
\cite{li96} and on various two-dimensional models \cite{li96,li98}.

\begin{figure}
%\vspace{6cm}
\centerline{\epsfxsize=8.5cm
\epsffile{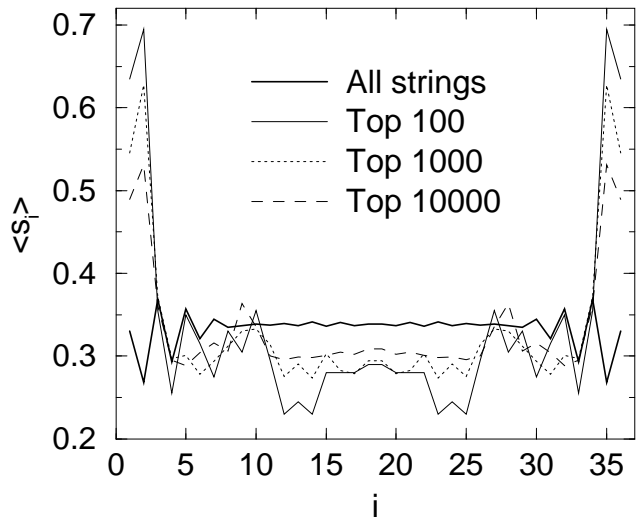}}
%\vspace {10cm}
\caption{The average $\langle s_i \rangle$ versus $i$. $s_i = 1$ for a
core site and $s_i = 0$ for a surface site. The average is taken over
all, top 100, top 1000, and top 10000 most designable structure strings,
respectively. (For any single string, the average of $s_i$ over
the string is 1/3.)}
\label{si}
\end{figure}

Also consistent with these previous works, there are noticeable
geometrical differences between the highly designable $4\times3\times3$
structures and the less designable ones. In Fig.~\ref{si}, we have plotted the
average structure string $\langle s_i \rangle$ for highly designable
structures, and for all structures. Since $s_i = 1$ for a core
site, and $s_i = 0$ for  a surface site, the ensemble average 
$\langle s_i \rangle$ gives the probability that the $i$th monomer
on the chain occupies a core site. It is seen in Fig.~\ref{si}
that for highly designable structures the first few monomers on 
the chain tend to occupy core sites, while there is no such tendency
for the average compact structure.  The average of $s_i$ 
for each chain is exactly 1/3 because every $4\times3\times3$ structure
has exactly 12 core sites ($s_i=1$) and 24 surface sites ($s_i=0$).
Therefore the tendency of the ends of highly 
designable structures to occupy core sites must 
be balanced by a tendency for the rest of the structure to occupy 
surface sites. This is also seen in Fig.~\ref{si} -- the central
third of the chain for highly designable structures has an increased
probability to occupy surface sites, on average.

\begin{figure}
%\vspace{6cm}
\centerline{\epsfxsize=8.5cm
\epsffile{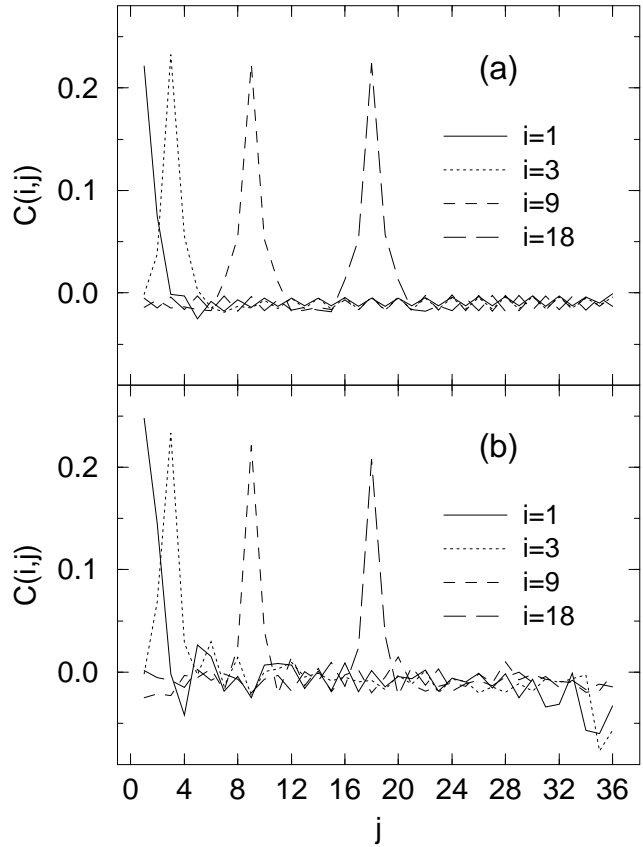}}
%\vspace {8cm}
\caption{Two-point correlation functions $C(i,j)$ of $s_i$, averaged
over (a) all structure strings and (b) over top 1000 most designable
strings.}
\label{corr}
\end{figure}

\begin{figure}
%\vspace{6cm}
\centerline{\epsfxsize=8.5cm
\epsffile{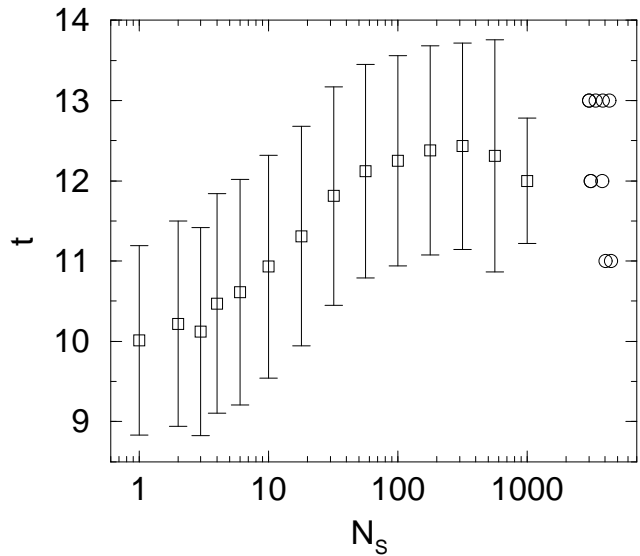}}
%\vspace {10cm}
\caption{The number of transitions $t$ between surface and core sites
versus designability. Structures with a set of selected $N_S$'s are shown.
Both the average (squares) and the rms deviations (error bars) for given
$N_S$'s are plotted. Also shown are the number of surface-core
transitions for ten topmost designable structures (circles).}
\label{s2c}
\end{figure}

In Fig.~\ref{corr} we have plotted the two-point correlation
function of structure strings, 
$C(i,j) = \langle s_i s_j \rangle - \langle s_i \rangle
\langle s_j \rangle$, averaged over highly designable structures
and over all structures. There is a clear correlation of site
types, with a range of roughly one monomer in either direction
along the chain. 
That is, if the $i$th monomer of a chain occupies a core site, there is an
enhanced probability for monomers $i \pm 1$ to occupy core sites.
This simply represents a general geometrical property of compact,
self-avoiding structures \cite{li98}.
The correlation length is slightly shorter for highly
designable structures (Fig.~\ref{corr}(b)), implying more frequent
transitions between surface and core sites. In Fig.~\ref{s2c}, we plot
the number of transitions $t$ between surface and core sites
versus designability $N_S$. For clarity, only structures with
selected $N_S$'s are included in the plot. We see a weak positive
correlation between $t$ and $N_S$, with a large variance for
a given $N_S$. A much stronger positive correlation between surface-core
transitions and designability was found in a two-dimensional $6\times6$
lattice model \cite{shih00,hell01}, possibly reflecting the larger more
compact core in the two-dimensional model.

\begin{figure}
%\vspace{6cm}
\centerline{\epsfxsize=8.5cm
\epsffile{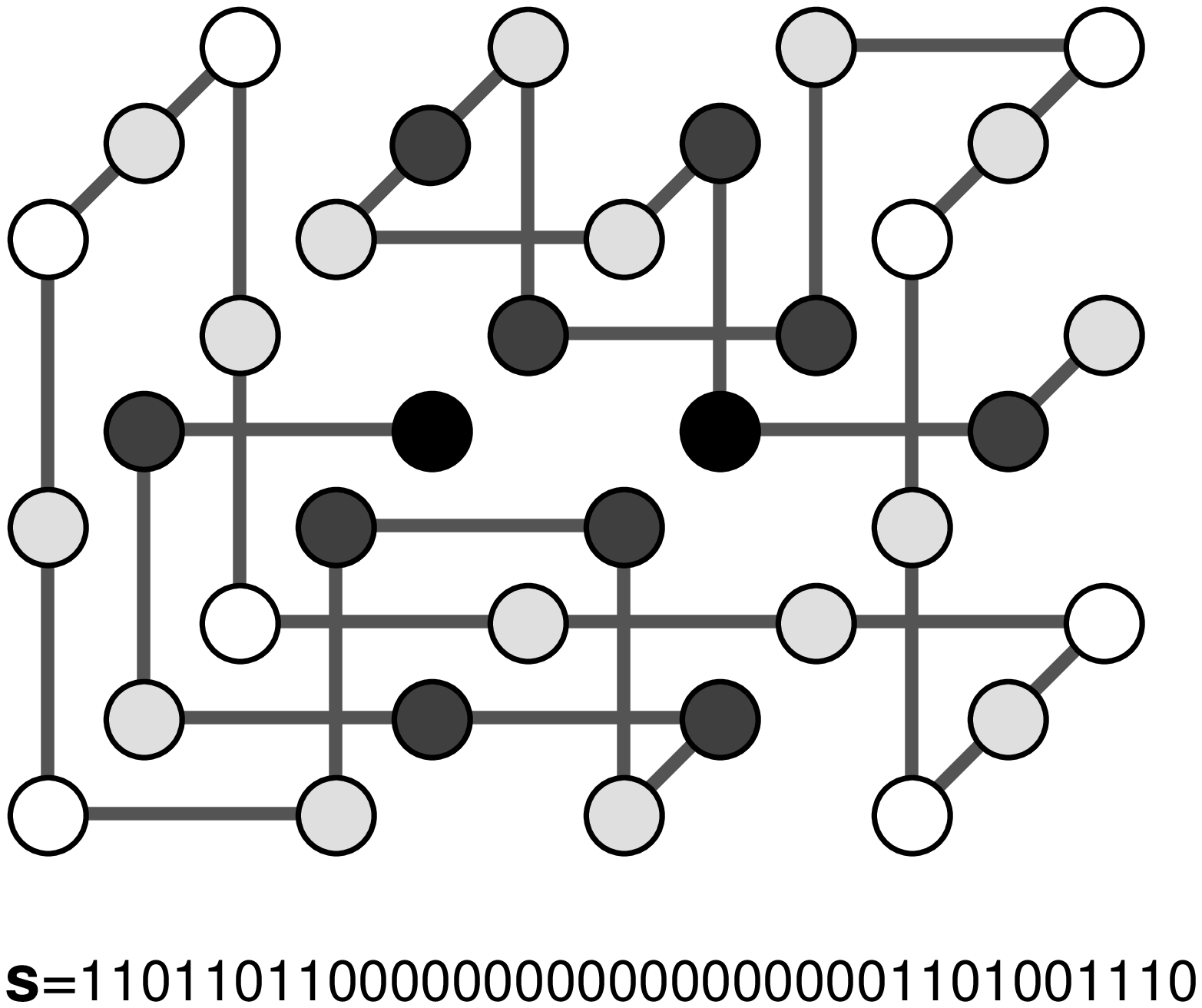}}
\vspace{1cm}
\caption{The topmost designable $4\times3\times3$ structure and its
structure string. (There is another topmost structure of the same geometrical 
conformation with the reversed chain direction.)}
\label{top}
\end{figure}

As an example, the topmost designable $4\times3\times3$ structure
is plotted in Fig.~\ref{top}. The geometry of this structure
is consistent with the results shown in Figs.~\ref{si}~and~\ref{corr}.
The 12 core sites are equally divided between the two ends of
the chain, with the center part of the chain, $i=9-26$, consisting
entirely of surface sites. Moreover, while the core sites tend
to cluster, the longest run of core sites is only three,
$i= 33-35$, consistent with the average correlation length shown
in Fig.~\ref{corr}. 

\begin{figure}
%\vspace{6cm}
\centerline{\epsfxsize=8.5cm
\epsffile{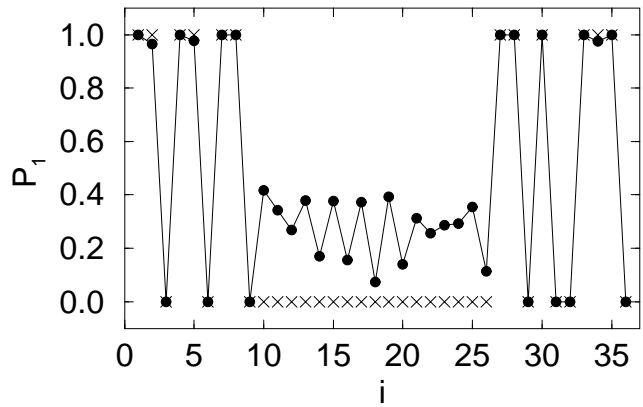}}
%\vspace{10cm}
\caption{The mutation pattern for the topmost designable structure,
shown in Fig.~(\ref{top}). The $\times$'s represent the structure
string, with 1's for core sites and 0's for surface sites.} 
\label{muta-top}
\end{figure}

The complete enumeration of sequences and structures allows us
to identify {\it all} sequences with a given structure as their
unique ground state. We can therefore analyze the statistical 
properties of sequences that design a particular structure.
For example, in Fig.~\ref{muta-top}, we have plotted the probability
that a hydrophobic monomer occupies position $i$, averaged over all
4466 sequences that design the structure in 
Fig.~\ref{top}. Fig.~\ref{muta-top} therefore represents
the complete mutation pattern of the topmost designable structure. 
The 12 core sites are easily identified
since the probability of a hydrophobic monomer at these sites
is nearly one. That is, nearly all of the 4466 sequences that
design this structure have 1's at these 12 positions. 
Similarly, the first three surface sites at the beginning
of the chain and the last four surface sites at the end of the chain 
are always occupied by polar monomers. Interestingly, the
monomers in the central parts of the chains, $i=10-26$, have a
roughly 1/3 chance of being hydrophobic even though all of these 
are surface sites. The mutation pattern therefore has the nontrivial 
feature that the monomers at some sites are critical for the stability
of the ground state while the monomers at other sites are freely mutable. 

\section{DISCUSSION}

The existence of highly designable structures emerged from a study
of $3\times3\times3$ and $6\times6$ HP lattice protein models with
interaction energies that included both solvation and segregation
components\cite{li96}.  A later study verified the existence
of highly designable structures for HP lattice models in two
dimensions including only solvation energies \cite{li98}.
Moreover, in two dimensions, there was little change in the qualitative
behavior of designability with increasing structure size,
suggesting that highly designable structures persist up to realistic protein
chain lengths. Within the $3\times3\times3$ solvation model, however, only
120 ($0.1\%$) out of the 103,346 total structures have nonzero designability.
(These 120 (60 lattice conformations) stand out as highly designable
structures with the largest
average gaps in the solvation plus segregation model of Li 
{\it et al.} \cite{li96}.) It has remained unclear whether the solvation
model in three dimensions can produce a significant fraction of
highly designable structures at realistic protein chain lengths. 
The current study, extending the solvation model up to
$4\times3\times3$ structures, offers strong evidence that
the existence of highly designable structures is a general
feature of solvation models in three dimensions.

To understand the ubiquitous appearance of highly designable 
structures, it is helpful to review the geometrical interpretation
of designability for the solvation model\cite{li98}.  
To this end, the energy in Eq.~(\ref{sol}) is rewritten as 
\begin{equation}
E =  \frac{1}{2}\sum_{i=1}^N \Bigl[ |\sigma_i - s_i|
     - |\sigma_i| - |s_i| \Bigr].
\label{dist}
\end{equation}
The last term is constant for a given sequence, and 
the second-to-last term is constant for all compact structures.
Therefore, the ground state for a given sequence is 
determined by the first term alone, which is one-half the Hamming distance
between the sequence string and the structure string.
Simply put, the structure nearest to a sequence is its ground state.
The designability of a structure is thus equal to the number of
sequences that lie closer to it than to any other structure. 

\begin{figure}
%\vspace{6cm}
\centerline{\epsfxsize=8.5cm
\epsffile{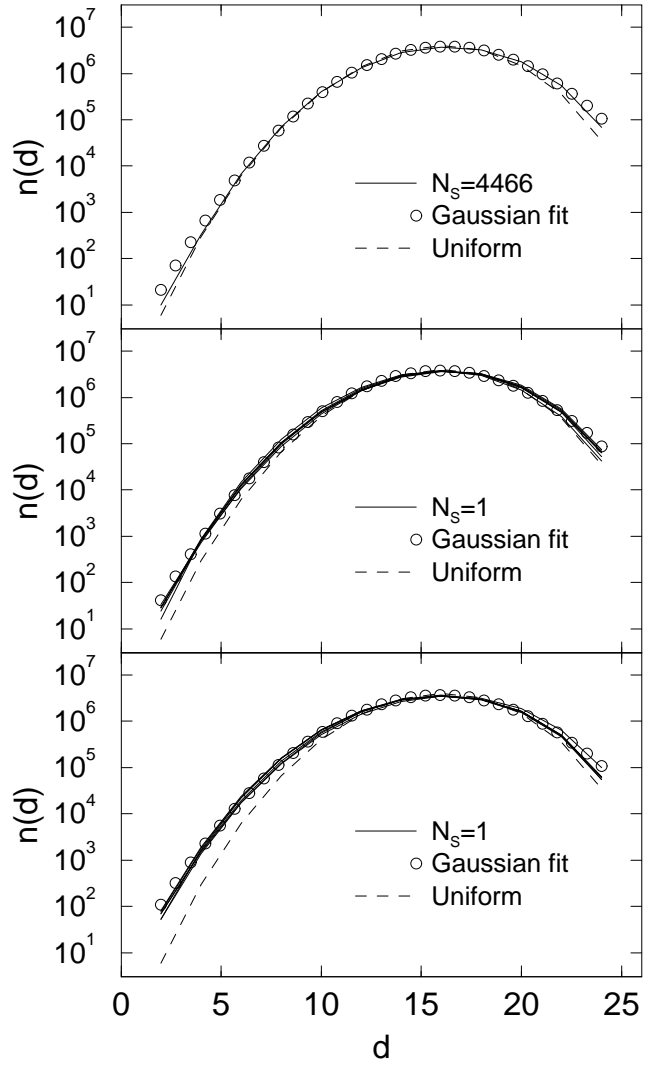}}
%\vspace{6cm}
\caption{Number of structure strings $n(d)$ at Hamming distance $d$ from
a given structure string.
(a) For the top structure;
(b) For 6 structures with $N_S=100$;
(c) For 6 structures with $N_S=1$.
In each case, the circles are a Gaussian distribution with the same mean
and variance. The dotted line is the expected $n(d)$ if the strings were
uniformly distributed.}
\label{nd}
\end{figure}

This geometrical interpretation suggests that highly designable 
structures are those with few nearby competing structures.
To test this, we have plotted in Fig.~\ref{nd} the number of neighboring
structures, as a function of Hamming distance between structure strings,
for the topmost designable structure and for structures of intermediate
($N_S =100$) and low ($N_S=1$) designability. It is seen that high
designability implies a reduced number of neighbors, and this correlation
persists out to distance 10 (structures whose strings differ by interchange
of five surface and core sites). For comparison, we have also plotted in
Fig.~\ref{nd} the expected $n(d)$ if the $14,062,236$ structure strings were
uniformly distributed on the hyperplane given by the constraints 
$\sum s_i=12$ and $\sum (-1)^i\times s_i=0$. The second constraint comes
from the fact that the $4\times3\times3$ cubic lattice is a bipartite
lattice, so that for any structure there are six core sites with $i$ even
and
six core sites with $i$ odd. The total number of points in the hyperplane
is $(C_{18}^6)^2=344,622,096$, where $C_n^m=n!/m!(n-m)!$. The number of
points in the plane at a Hamming distance $d=2 \times l$ from a given point
is simply: $N(d)=\sum_{k=0}^l C_6^k C_{12}^k C_6^{l-k} C_{12}^{l-k}$,
with $C_n^m=0$ if $m>n$. The expected $n(d)$ is then $\rho \times N(d)$,
where $\rho=14,062,236/344,622,096 \approx 0.04$ is the average density of
real structure strings in the hyperplane. We see that the structure strings
are more clustered than the uniform distribution -- the lower the
designability, the more clustered they are.

\begin{figure}
%\vspace{6cm}
\centerline{\epsfxsize=8.5cm
\epsffile{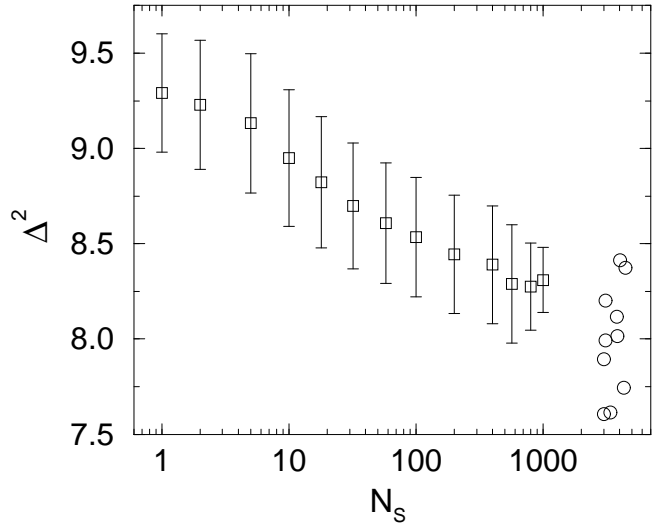}}
%\vspace{10cm}
\caption{The second moment $\Delta^2 (\equiv \langle d^2\rangle -\langle
d\rangle^2)$ of distribution of neighbor distances $n(d)$ versus the
designability. The error bars indicate the rms deviations of $\Delta^2$
for given $N_S$. The circles are $\Delta^2$ for the ten topmost designable
structures.}
\label{nsmom2}
\end{figure}

A scarcity of near neighbors 
corresponds to a narrow width of the distribution of neighbor distances,
since every structure has the same total number of neighbors. 
We have therefore plotted in Fig.~\ref{nsmom2} the width of the distribution
of neighbor distances over the entire range of designability.
The width falls smoothly with increasing designability, indicating
a general correlation between high designability of a structure 
and a scarcity of nearby competing structures.

The existence of highly designable structures can therefore
be viewed as a geometrical property of the space of structure
strings. For the two dimensional solvation model, it has been 
stressed \cite{li98} that the most highly
designable structures fall in regions of low density
in the space of structure strings. In this sense, highly 
designable structures have ``atypical" patterns of surface and
core sites. The current work extends this conclusion to 
a three dimensional solvation model with chain lengths
($N=36$) approaching those of real proteins.

In summary, we have employed a fast tree-search algorithm to find
the ground states of all sequences for a $4\times3\times3$ lattice HP
model of proteins. The results confirm the existence of highly designable
structures in a three-dimensional solvation model with a surface-to-core
ratio of 2-to-1, close to the values for small natural proteins. 
Highly designable structures are found to differ geometrically from other
structures. Interestingly, structures are found to have 
nontrivial mutation patterns with some sites strictly
conserved and others mutable. The fast tree-search algorithm
is particularly well suited to lattice HP solvation models,
and we hope our detailed description of the method will prove useful.

\newpage

%%%%%%%%%%%%%%%%%%FIGURE 1%%%%%%%%%%%%%%%%%%%%
%%%%%%%%%%%%%%%%%%%%%%%%%%%%%%%%%%%%%%%%%%%%%%

\newpage

%%%%%%%%%%%%%%%%%%FIGURE 2%%%%%%%%%%%%%%%%%%%%
%%%%%%%%%%%%%%%%%%%%%%%%%%%%%%%%%%%%%%%%%%%%%%

\newpage

%%%%%%%%%%%%%%%%%%FIGURE 3%%%%%%%%%%%%%%%%%%%%
%%%%%%%%%%%%%%%%%%%%%%%%%%%%%%%%%%%%%%%%%%%%%%

%%%%%%%%%%%%%%%%%%FIGURE 4%%%%%%%%%%%%%%%%%%%%
%%%%%%%%%%%%%%%%%%%%%%%%%%%%%%%%%%%%%%%%%%%%%%

%%%%%%%%%%%%%%%%%%FIGURE 5%%%%%%%%%%%%%%%%%%%%
%%%%%%%%%%%%%%%%%%%%%%%%%%%%%%%%%%%%%%%%%%%%%%

%%%%%%%%%%%%%%%%%%FIGURE 6%%%%%%%%%%%%%%%%%%%%
%%%%%%%%%%%%%%%%%%%%%%%%%%%%%%%%%%%%%%%%%%%%%%

%%%%%%%%%%%%%%%%%%FIGURE 7%%%%%%%%%%%%%%%%%%%%
%%%%%%%%%%%%%%%%%%%%%%%%%%%%%%%%%%%%%%%%%%%%%%

%%%%%%%%%%%%%%%%%%FIGURE 8%%%%%%%%%%%%%%%%%%%%
%%%%%%%%%%%%%%%%%%%%%%%%%%%%%%%%%%%%%%%%%%%%%%

%%%%%%%%%%%%%%%%%%FIGURE 9%%%%%%%%%%%%%%%%%%%%
%%%%%%%%%%%%%%%%%%%%%%%%%%%%%%%%%%%%%%%%%%%%%%

%%%%%%%%%%%%%%%%%%FIGURE 10%%%%%%%%%%%%%%%%%%%%
%%%%%%%%%%%%%%%%%%%%%%%%%%%%%%%%%%%%%%%%%%%%%%

\end{document}